\documentclass[11pt,preprintnumbers,aps,amssymb,nofootinbib,amsmath,superscriptaddress,notitlepage,prd]
{revtex4-1}
\usepackage{epsfig,epsf}
\usepackage{comment}
\usepackage{bm} 
\usepackage{color} 
\usepackage{slashed} 
\usepackage{relsize}	
\usepackage{soul} 
\usepackage{hyperref}
\usepackage{tensor} 
\newcommand{\beq}{\begin{equation}}
\newcommand{\beql}[1]{\begin{equation}\label{#1}}
\newcommand{\eeq}{\end{equation}}
\def\bal#1\gal{\begin{align}#1\end{align}}
\newcommand{\ball}[1]{\bal\label{#1}}
\newcommand{\eq}[1]{(\ref{#1})}
\newcommand{\fig}[1]{Fig.~\ref{#1}}

\DeclareMathOperator{\im}{\mathrm{Im}}

\renewcommand{\b}[1]{{\bm #1}} 

\newcommand{\aver}[1]{\left\langle #1 \right\rangle}
\newcommand{\arctanh}{\text{arctanh\,}}
\usepackage{physics}

\begin{document}

\title{Bremsstrahlung in chiral medium: anomalous magnetic contribution to the Bethe-Heitler formula}

\author{Jeremy Hansen}

\author{Kirill Tuchin}

\affiliation{
Department of Physics and Astronomy, Iowa State University, Ames, Iowa, 50011, USA}

\date{\today}

\begin{abstract}
 
We investigate photon bremsstrahlung  in chiral media. The chiral medium response to the magnetic field is described by the chiral magnetic current $\b j= b_0 \b B$, where $b_0$ is the chiral magnetic conductivity. This current modifies the photon dispersion relation producing a resonance  in the scattering amplitude. We show that the resonant contribution is proportional to the magnetic moment $\mu$ of the target nucleus. We analytically compute the corresponding cross section.  We argue that the anomalous contribution is enhanced by a factor $b_0^2/\Gamma^2$, where $\Gamma$ is a width of the  resonance related to the chiral magnetic instability of the electromagnetic field. The most conspicuous feature of the anomalous contribution to the photon spectrum is the emergence of the knee-like structure at photon energies proportional to $b_0$.  We argue that the phenomenological significance of the anomalous terms depends on the magnitude of the ratio of $b_0$ to the projectile fermion mass. 

\end{abstract}

\maketitle

\section{Introduction}\label{sec:a}

Chiral media are of great interest in a variety of diverse physics fields \cite{Zhitnitsky:2012ej,Kharzeev:2007tn,Kharzeev:2015znc,Li:2014bha,Gorbar:2018nmg,Marsh:2015xka,Klinkhamer:2004hg}. Examples include the quark-gluon plasma, Weyl and Dirac semimetals and the axion---one of the main dark matter candidates. Their unusual properties are due to the topological charge induced in them by the chiral anomaly \cite{Adler:1969gk,Bell:1969ts}.  In particular, coupling of the electromagnetic field to the topological charge  is  described by adding to the QED Lagrangian the $P$ and $CP$-odd  term  \cite{Fujikawa:2004cx} 
\ball{a1}
\mathcal{L}_A =-\frac{c_A}{4}\theta F_{\mu\nu}\tilde F^{\mu\nu}\,,
\gal
where $c_A$ is the QED anomaly coefficient. In quark-gluon plasma the dimensionless pseudoscalar field $\theta$ is sourced by the topological charge density
$q= \frac{g^2}{32\pi^2}G_{\mu\nu}^a\tilde G^{a\mu\nu}$.
As a result \eq{a1} cannot be rewritten as a total derivative and removed from the Lagrangian. Instead, it appears in the modified Maxwell equations as the derivative of $\partial^\mu\theta$. In Weyl semimetals the spatial components of $\partial^\mu\theta$ are proportional to the distance between the Weyl nodes in the momentum space.  
In many systems $\theta$ is believed to be a slowly varying function of coordinates and time \cite{Kharzeev:2015znc}. This is the approximation also assumed in this work. In particular, we treat the first derivative $\partial\theta$ as constant and adopt a fairly standard notation  $b^\mu= (b_0,-\b b)=c_A\partial^\mu \theta= c_A(\dot \theta, -\b \nabla \theta)$; $b_0$ is also known as the chiral conductivity $\sigma_\chi$  \cite{Kharzeev:2009pj,Fukushima:2008xe}. To further simplify the analysis we constrain ourselves to the spatially homogenous systems  with $|\b b|\ll b_0$. These are relevant in nuclear physics  and astrophysics \cite{Zhitnitsky:2014ria,Zhitnitsky:2014dra}. 
In this paper we focus on the electromagnetically interacting systems, but its methodology can be easily adapted for the strongly interacting systems as well.

\begin{figure}[ht]
      \includegraphics[height=4cm]{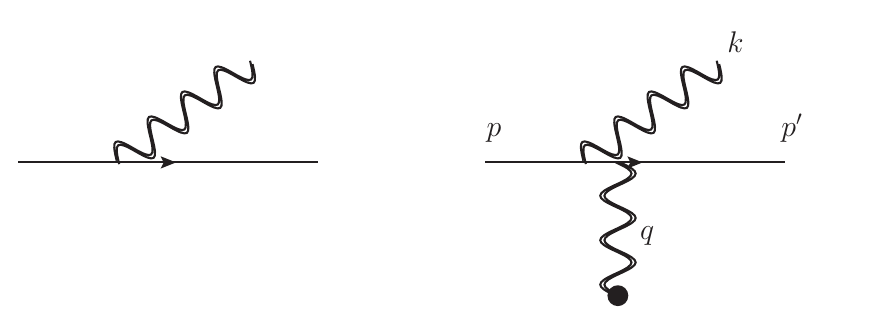} 
  \caption{Diagrams contributing to the collisional (left) and radiative (right) energy loss in the chiral medium at the leading order in $\alpha$. The left diagram includes the Cherenkov radiation. Double wavy lines indicate excitation of the electromagnetic field at finite $b$. The diagram with the photon emitted from the outgoing fermion leg is not shown.  }
\label{diagrams2}
\end{figure}

One can study properties of a novel material/medium by shooting fast particles through it. This paper studies how the chiral properties of a material are reflected in the photon spectrum emitted by a charge fermion. The two processes that determine the photon spectrum are the medium polarization by the electromagnetic fields of the particle and the photon radiation. The former induces the collisional energy loss which includes the Cherenkov radiation \cite{Fermi:1940zz,Tuchin:2018mte}, while the latter is responsible for the radiative energy loss. 
The corresponding leading order diagrams are shown in \fig{diagrams2}.

The collisional energy loss in chiral medium is dominated by the chiral Cherenkov radiation, which in many aspects is very different from the conventional Cherenkov radiation due to a peculiar dependence of the photon dispersion relation on $b$.  We  have discussed this recently in \cite{Hansen:2020irw} in the framework of  the Fermi's model \cite{Fermi:1940zz}. In short, 
the non-anomalous and anomalous/chiral Cherenkov radiation are both electromagnetic excitations of the medium induced by the moving charge though they emerge under different conditions: the former one when the particle velocity is larger than the phase velocity of light, whereas the latter when the chiral anomaly is present regardless of velocity. Unlike the non-anomalous Cherenkov radiation whose power decreases with energy (together with the total collisional loss), the power of the chiral Cherenkov radiation increases with energy and thus is very much relevant at high energies. The full quantum expression for the radiation power was derived in \cite{Tuchin:2018sqe}.  Similar results in different contexts were reported in \cite{Lehnert:2004hq,Lehnert:2004be,Klinkhamer:2004hg}. Photon radiation by a plasma at finite axial chemical potential, which also induces the chiral magnetic current, was investigated in \cite{Carignano:2018thu,Carignano:2019zsh,Carignano:2021mrn}.

The goal of this paper is to investigate the contribution of the chiral anomaly to photon production cross section in charged fermion scattering off a heavy nucleus. This process is responsible for the radiative energy loss which is  the dominant energy loss mechanism at high energy \cite{Peigne:2008wu}. The chiral anomaly affects the virtual photon propagator and the radiated photon wave function. The photon propagator  reads in the Feynman gauge  \cite{Carroll:1989vb,Lehnert:2004hq}\footnote{The gauge dependence of the propagator is discussed in Appendix~\ref{ap1}.}:  
\ball{a5}
D_{\mu\nu}(q)= -i \frac{q^2 g_{\mu\nu}+i\epsilon_{\mu\nu\rho \sigma}b^\rho q^\sigma+b_\mu b_\nu}{q^4+b^2 q^2-(b\cdot q)^2}\,.
\gal
In homogeneous chiral matter with $\b b=0$, $b_0\neq 0$, the components of the propagator \eq{a5} read in the static limit  $D_{\mu\nu}(\b q)= \lim_{q^0\to 0} D_{\mu\nu}(q)$ \cite{Qiu:2016hzd}
\begin{subequations}\label{b12}
\bal
&D_{00}(\b q)= \frac{i}{\b q^2}\,,\label{b12a}\\
&D_{0i}(\b q)= D_{0i}(\b q)= 0\,,\label{b12b}\\
&D_{ij}(\b q)=-\frac{i\delta_{ij}}{\b q^2-b_0^2}-\frac{\epsilon_{ijk}q^k}{b_0(\b q^2-b_0^2)}+\frac{\epsilon_{ijk}q^k}{b_0\b q^2}\,.\label{b12c}
\gal
\end{subequations}
The pole at $\b q^2=b_0^2$ is related to the chiral magnetic instability of electromagnetic field as discussed in Sec.~\ref{sec:instab}.  In the scattering amplitude it corresponds to the $t$-channel resonance which enhances the scattering cross sections in the chiral medium  \cite{Tuchin:2020gtz}. To simplify the calculation we restrict ourselves to the photon spectrum at $\omega\gg b_0$. In this region, photon is approximately timelike up to the corrections of order $b_0^2/\omega^2$. In other words, we neglect the anomaly in the photon wave function and thereby isolate the contribution of the resonance in the propagator.  

 In summary, we consider the following setup. A charged fermion of high energy $\epsilon$ radiates a photon of energy $\omega$ as it scatters on a heavy (static) ion immersed into a chiral medium. The potential that the ion creates is modified in the infrared by the anomaly scale $b_0$ which is assumed to be much smaller than $\omega$, but  much larger than the Debye mass $eT$ in the medium. In this regime the bremsstrahlung cross section is enhanced by the $t$-channel resonance at the momentum transfer $\b q^2=b_0^2$. The softest scale in this scattering problem (apart from the Debye mass) is the resonance cutoff $\Gamma$ which is inversely proportional to the duration of the inverse cascade that drives the chiral magnetic instability. In the next section we perform a detailed analytical calculation of the photon production cross section. Our main result is Eq.~\eq{j3} which presents this cross section in the ultra-relativistic limit. We discuss the obtained result in Sec~\ref{sec:dis}.

\section{Coulomb and magnetic dipole bremsstrahlung}\label{sec:b}

\subsection{Scattering potential}

Following the original calculation of Bethe and Heitler \cite{Bethe:1934za}, we consider scattering of a charged fermion off a nucleus of mass $M$ and electric charge $eZ$. It will be seen in Sec.~\ref{sec:j} that the anomalous contribution is driven by the existence of momentum transfers such that $|\b q|\le b_0$. The nucleus recoil  is $q_0=|\b q|^2/(2M) < |\b q|b_0/(2M)$. 
Since for any realistic system $b_0\ll M$, the recoil can be safely neglected. Thus the electromagnetic field of the nucleus is approximately static. 

The potential induced by a stationary current $J^\nu(\b x)$ can be computed as  
\ball{b11}
A^\mu (\b x) = -i\int d^3x' D^{\mu\nu}(\b x-\b x')J_\nu(\b x')= -i\int \frac{d^3q}{(2\pi)^3} e^{i\b q\cdot \b x}D^{\mu\nu}(\b q) J_\nu (\b q)\,.
\gal
The current density of the static point source of charge $eZ$ is $J^\nu(\b x) =eZ \delta\indices{^\nu_0}\delta(\b x)$. It induces the Coulomb potential 
\ball{b15}
A^0(\b q)= eZ/\b q^2\,,\qquad \b A(\b q)=0\,
\gal
implying that the scattering cross section off the point charge is given by the Rutherford formula and is not affected by the anomaly (in the static limit).  

A non-trivial contribution comes about if the nucleus is in a state $\psi$ with a finite expectation value of the magnetic moment $\b \mu$. Indeed, the spin current associated with such a state is $\b\nabla \times \psi^* \b \mu \psi$. In the point particle limit the spin current can be written as  $\b J(\b x) = \b \nabla \times (\b \mu\delta(\b x))$. It represents the first non-vanishing  multipole moment of the vector potential. Altogether the electrical current of the nucleus is
\ball{b16}
J^0(\b x) =eZ\delta(\b x)\,,\qquad 
\b J(\b x) = \b \nabla \times (\b \mu\delta(\b x))\,,
\gal
which in momentum space reads
\ball{b17}
J^0(\b q)=eZ\,,\qquad \b J(\b q)= i\b q\times \b \mu\,.
\gal
According to \eq{b11} and  \eq{b12} it  produces the potential 
\begin{subequations}\label{b19}
\bal
A^\ell(\b q)&=-iD^{\ell i}(\b q)J_i(\b q)=  -\epsilon_{ijk}\mu^kq^j\left( -\frac{i\delta_{\ell i}}{\b q^2-b_0^2}-\frac{\epsilon_{\ell i r}q^r}{b_0(\b q^2-b_0^2)}+\frac{\epsilon_{\ell i r}q^r}{b_0\b q^2}\right)\label{b19a}\\
&
= -\frac{1}{\b q^2-b_0^2}\left[ i(\b \mu\times \b q)^\ell + \frac{b_0}{\b q^2}(\b \mu\cdot \b q q^\ell-\b q^2 \mu^\ell)\right]\,, \label{b19b}
\gal
\end{subequations}
while the time component is still given by the first equation of \eq{b15}. The potential satisfies the Coulomb gauge $\b q\cdot \b A=0$. Expression for the potential in the configuration space can be found in \cite{Tuchin:2020gtz}.

\begin{figure}[ht]
      \includegraphics[height=4cm]{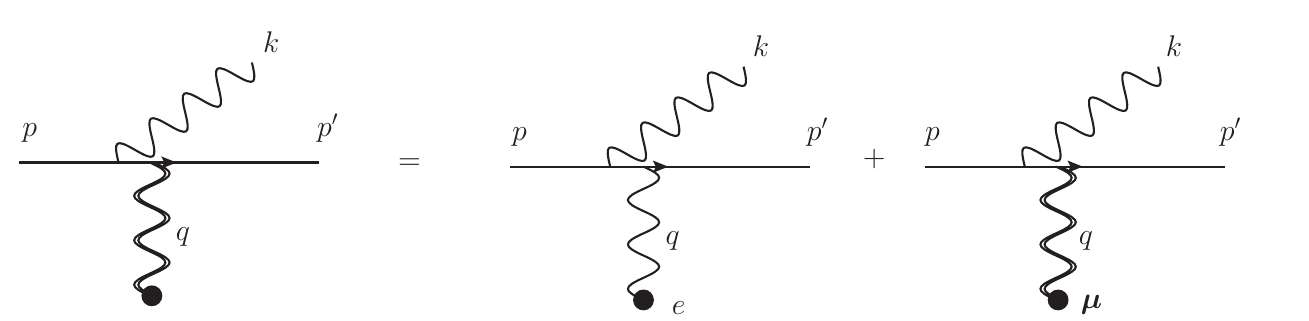} 
  \caption{Diagrams contributing to the Coulomb ($e$) and magnetic moment ($\mu$) terms in the bremsstrahlung cross section in the chiral medium at the leading order in $\alpha$. The outgoing photon is depicted by a single line, which is accurate up to terms of the order $b_0^2/\omega^2$.}
\label{diagrams}
\end{figure}

\subsection{Scattering cross section}

Quantization of the electromagnetic field at finite $b_0$ is similar to this procedure in vacuum except that the photon polarization must be circular, see Appendix~\ref{sec:app2}. As explained in Introduction, in this work we consider photons with $\omega\gg b_0$ that allows us to neglect the chiral Cherenkov effect and concentrate on the bremssrahlung per se. The differential cross section is then given by
\ball{c1}
    d\sigma=\frac{1}{2}\sum_{ss'\lambda}\left|\mathcal{M}\right|^2
    \frac{1}{8(2\pi)^5}\frac{\omega |\b p'|}{|\b p|} d\Omega_{\bm{k}} d\Omega' d\omega
\gal
Where $p=(\epsilon,\bm{p})$,  $p'=(\epsilon',\b p')$ and $k=(\omega,\bm{k})$ are the incoming, outgoing and emitted photon four momenta respectively. The corresponding equations of motion imply that $p^2=p'^2=m^2$ and $k^2=-\lambda b_0|\b k|$, where $\lambda=\pm 1$ is the photon polarization. As mentioned before, we assume that $\omega\gg b_0$, so that $k^2\approx 0$.
The matrix element $\mathcal{M}$ can be written as a sum corresponding to two Feynman diagrams where the photon is emitted before and after the insertion of the external field $A(\bm{q})$, where $q=p'-p+k$:
\ball{c3}
    \mathcal{M}=e^2 \overline{u}(p')\left(\slashed{e}_{k\lambda}^*\frac{ \slashed{p}'+\slashed{k}+m}{(p'+k)^2-m^2}\slashed{A}(\b q)+\slashed{A} (\bm{q})\frac{ \slashed{p}-\slashed{k}+m}{(p-k)^2-m^2}\slashed{e}_{k\lambda}^*\right)u(p)\,,
\gal
where $e_{k\lambda}^\mu= (0,\b e)$ is  the circular photon polarization vector. 
Plugging in Eqs.~\eq{b15},\eq{b19b} into \eq{c3} and averaging over directions of $\b\mu$ using  $\aver{\mu_i}=0$, $\aver{\mu_i\mu_j}=\frac{\mu^2}{3}\delta_{ij}$ yields
 \ball{c7}
     \abs{\mathcal{M}}^2=\abs{\mathcal{M}_e}^2+\abs{\mathcal{M}_\mu}^2
\gal
where $|\mathcal{M}_{e}|^2$ is proportional to $e^2Z^2$ and describes scattering off the Coulomb field \eq{b15}, see \fig{diagrams}. It does not depend on $b_0$ and ultimately leads to the Bethe-Heitler formula when substituted into \eq{c1}. The second term, $|\mathcal{M}_\mu|^2$ is proportional to $\mu^2$ and carries all the information about the anomaly. We will focus on it from now on. Averaging over the magnetic moment directions in \eq{b19b} gives
\ball{c9}
\aver{A^i(\b q)A^{j*}(\b q)}=\frac{\mu^2}{3(\b q^2-b_0^2)^2}\left[ \left(\delta^{ij}-\frac{q^iq^j}{\b q^2}\right)(\b q^2+b_0^2)-2ib_0\epsilon^{ijk}q_k\right]\,.
\gal
Its contribution to the cross section has the form $\mathcal{M}_i\mathcal{M}_j^*\aver{A^iA^{j*}}$. When summed over the particle helicities $\mathcal{M}_i\mathcal{M}_j^*$ is symmetric with respect to swapping the indices $i$ and $j$. This implies that the second term in the square brackets of \eq{c9} cancels out.\footnote{We note however, that this term  contributes to the partial cross sections for the helicity states.} 

It is convenient to express the cross section in terms of $q^2=-\b q^2$, $\omega$ and the parameters $\kappa$ and $\kappa'$ defined as
\ball{c10a}
\kappa=k\cdot p/\omega= \epsilon-\bm{k}\cdot \bm{p}/\omega\,,\qquad \kappa'=k\cdot p'/\omega= \epsilon'-\bm{k}\cdot \bm{p'}/\omega\,.
\gal
Dotting the momentum conservation condition $p+q-k-p'=0$ with $p$, $q$, $k$ and $p'$ and solving the resulting four equations we obtain
\bal
q\cdot p = -q^2/2+\kappa' \omega\,,
\qquad 
q\cdot p' = q^2/2+\kappa \omega\,,\label{c10b}\\
p\cdot p'= m^2-\omega(\kappa-\kappa')-\frac{q^2}{2}\,,\qquad k\cdot q= \omega(\kappa'-\kappa)\,. \label{c10c}
\gal
Summing over the outgoing fermion spin and photon polarization and averaging over the incident fermion spin we derive
\ball{fullmu}
\frac{1}{2}\sum_{ss'\lambda}\abs{\mathcal{M}_{\mu}}^2=&\frac{4 e^4 \mu ^2 (b_0^2+\b q^2) }{3 \kappa^2 \kappa'^2  \omega ^2 (\b q^2-b_0^2)^2}\bigg\{
2\kappa\kappa'\omega^2\Big[ \left(\kappa+\kappa'\right)^2 +\bm{q}^2\Big]
\nonumber\\
&
+\sum_{\lambda}\Big [\bm q^2\abs{\kappa   \b p'\cdot \b e -\kappa' \b p\cdot \b e}^2+
4\kappa^2 \bm{p}^2 \abs{   \b p'\cdot \b e}^2+4\kappa '^2 \b p'^2\abs{ \b p\cdot \b e}^2
\nonumber\\
&
-8 \kappa \kappa' \left(  \bm{p}\cdot \b e\right) \left(  \b p'\cdot \b e^*\right) \left( \epsilon  \epsilon'- m^2\right)
\Big ] \bigg\}\,,
\gal
The polarization sums are computed using \eq{ap23}:
\begin{subequations}\label{d4}
\bal 
&\sum_\lambda \abs{\b e \cdot\bm{ p}}^2=\b p^2-\frac{(\b p\cdot \b k)^2}{\b k^2}
=\b p^2-(\epsilon-\kappa)^2=2\epsilon\kappa-m^2-\kappa^2\,,\label{d1}\\
&  \sum_\lambda \abs{\b e\cdot \b p'}^2=\b p'^2-\frac{(\b p'\cdot \b k)^2}{\b k^2}
=2\epsilon'\kappa'-m^2-\kappa'^2\,,\label{d2}\\
&   \sum_\lambda (\b e\cdot\b  p)(\b e^*\cdot \b p')=\b p \cdot \b p'-\frac{(\b p\cdot \b k) (\b p'\cdot \b k)}{\b k^2}
=\kappa'\epsilon'+\kappa\epsilon-m^2-\kappa\kappa'+\frac{q^2}{2}\,,
\label{d3}
\gal    
 \end{subequations}
 where \eq{c10a},\eq{c10b},\eq{c10c} where used. 
 Employing \eq{d4} in \eq{fullmu} and substituting into \eq{c1} we obtain the differential cross section 
 \ball{d6}
d\sigma_\mu=&\frac{m^2 e^4 \mu ^2 (b_0^2+\b q^2) }{6(2\pi)^5 \omega (\b q^2-b_0^2)^2}\frac{ |\b p'|}{|\b p|} d\Omega_{\bm{k}} d\Omega' d\omega
\nonumber\\
&\times\Big\{\frac{\bm q^2}{m^2 \kappa \kappa'}(2\epsilon^2+2\epsilon'^2-4m^2+\b q^2)+(4m^2-\b q^2)\left(\frac{1}{\kappa}-\frac{1}{\kappa'}\right)^2-4\left(\frac{\epsilon}{\kappa'}-\frac{\epsilon'}{\kappa}\right)^2
\nonumber\\
&+\frac{2\omega}{m^2}(4m^2-\b q^2)\left(\frac{1}{\kappa'}-\frac{1}{\kappa}\right)+\frac{2\omega^2}{m^2}\left(\frac{\kappa'}{\kappa}+\frac{\kappa}{\kappa'}\right) \Big\}\,.
\gal

\subsection{Regularization of the resonance}\label{sec:instab}

In a homogeneous medium ($\b b=0$) the photon propagator \eq{a5} has a resonance at $q^2=M^2=-\lambda b_0 |\b q|$. This resonant behavior can be regulated in the usual way by taking account of the finite resonance width:
\ball{f1}
\frac{1}{q^4+b^2q^2-(b\cdot q)^2}\to \frac{1}{q^4+b^2q^2-(b\cdot q)^2+i q^2\Gamma^2}\,,
\gal
so that the denominator now vanishes at $q^2=M^2-i\Gamma^2/2$. 
It is noteworthy that the photon propagator exhibits both the $s$-channel ($q^2>0$) and the $t$-channel ($q^2<0$) resonances. This happens because $M^2$ can be positive or negative depending on the photon polarization. The same parameter $\Gamma$ regulates both channels. We note  
that $\Gamma$ in the $s$-channel has a transparent physical meaning. Namely, it is related to the photon decay width $W$ as $\Gamma^2= M W=\sqrt{|b_0||\b q|}W\approx b_0 W$. 

The resonant behavior is closely related to the chiral magnetic instability of electromagnetic field in chiral medium which is driven by the modes with $\im q^0>0$ see e.g.\
\cite{Joyce:1997uy,Boyarsky:2011uy,Hirono:2015rla,Xia:2016any,Kaplan:2016drz,Kharzeev:2013ffa,Khaidukov:2013sja,Avdoshkin:2014gpa,Akamatsu:2013pjd,Kirilin:2013fqa,Tuchin:2014iua,Dvornikov:2014uza,Buividovich:2015jfa,Sigl:2015xva,Kirilin:2017tdh,Tuchin:2017vwb}. From the dispersion relation 
$(q^0)^2=\b q^2-\lambda b_0 |\b q|$
it is evident that these modes have $\b q^2< b_0^2$. In the limit of small $q^0$ there is only one unstable mode $|\b q|=b_0$. The instability is eventually tamed by the chirality flow between the magnetic field and the medium, which induces the time-dependence of $b_0$. It is reasonable then to estimate $W$ as the inverse of the chirality transfer time $W\sim \alpha^2m^2/T$ \cite{Boyarsky:2011uy} ($T$ is the tempearture) which is the softest scale in the problem. Another contribution to $W$ arises from inhomogeneity of the axial charge distribution which induces spontaneous transitions between the eigenstates of the curl operator (also known as the Chandrasekhar-Kendall states) \cite{Tuchin:2016qww}. 

In our calculation, the denominators of \eq{f1} appears multiplied by its complex conjugate. Taking advantage of the fact that $q^0=0$ we can cast this product in a form convenient for further analysis:
\begin{subequations}\label{f6}
\ball{f3}
    \Re\Big[\frac{\Gamma^2-2ib_0^2}{\Gamma^2(\b q^2-b_0^2-i\Gamma^2)}\big]&=\Re\Big[\frac{\Gamma^2(\b q^2-b_0^2)+2b_0^2\Gamma^2+i(\Gamma^4-2b_0(\b q^2-b_0^2))}{\Gamma^2((\b q^2-b_0^2)^2+\Gamma^4)}\big]
    \nonumber\\
    &=\frac{(\b q^2+b_0^2)}{(\b q^2-b_0^2)^2+\Gamma^4}\,.
\gal
Similarly,
\bal
    \frac{\b q^2(\b q^2+b_0^2)}{(\b q^2-b_0^2)^2+\Gamma^4}&=\Re\Big[\frac{3b_0^2\Gamma^2+(\Gamma^4-2b_0^4)i}{\Gamma^2(\b q^2-b_0^2-i\Gamma^2)}+1\big]\,,\label{f4}\\
    \frac{\b q^4(\b q^2+b_0^2)}{(\b q^2-b_0^2)^2+\Gamma^4}&=\Re\Big[\frac{\Gamma^2(5 b_0^4-\Gamma^4)+2b_0^2(2\Gamma^4-b_0^4)i}{\Gamma^2(\b q^2-b_0^2-i\Gamma^2)}+(q^2+3b_0^2)\big]\,.\label{f5}
\gal
\end{subequations}
Using Eqs.~\eq{f6} in \eq{d6} furnishes the final expression for the regulated cross section
\ball{imsig1}
\frac{d\sigma_\mu}{d\Omega_{\bm{k}} d\Omega' d\omega}=\frac{m^2 e^4 \mu ^2  }{6(2\pi)^5 \omega\Gamma^2 }\frac{ |\b p'|}{|\b p|} \Re\Bigg\{\frac{1}{m^2 \kappa \kappa'}\left[\frac{\Gamma^2(5 b_0^4-\Gamma^4)+2b_0^2(2\Gamma^4-b_0^4)i}{Q}+(q^2+3b_0^2)\Gamma^2\right]\nonumber\\
+\left[\frac{3b_0^2\Gamma^2+(\Gamma^4-2b_0^4)i}{Q}+\Gamma^2\right]
\left[\frac{2\epsilon^2+2\epsilon'^2-4m^2}{m^2 \kappa \kappa'}-\left(\frac{1}{\kappa}-\frac{1}{\kappa'}\right)^2-\frac{2\omega}{m^2}\left(\frac{1}{\kappa'}-\frac{1}{\kappa}\right)\right]\nonumber\\
+\left(\frac{\Gamma^2-2b_0^2i}{Q}\right)\left[4m^2\left(\frac{1}{\kappa}-\frac{1}{\kappa'}\right)^2+8\omega\left(\frac{1}{\kappa'}-\frac{1}{\kappa}\right)-4\left(\frac{\epsilon}{\kappa'}-\frac{\epsilon'}{\kappa}\right)^2+\frac{2\omega^2}{m^2}\left(\frac{\kappa'}{\kappa}+\frac{\kappa}{\kappa'}\right)\right] \Bigg\}\,,
\gal
where $Q=\b q^2-b_0^2-i\Gamma^2$.


The integral over the final fermion directions $d\Omega'$ can be done explicitely by introducing the Feynman parameter $x$ \cite{Gluckstern:1953zz}. For example,
\ball{i11}
    I_{1,1}&\equiv\int\frac{\epsilon^4d\Omega d\Omega'}{(2\pi)^2\kappa\kappa'Q}=\int\frac{\epsilon^4d\Omega}{(2\pi)^2\kappa((\b p-\b k)^2+\b p^{'2})\epsilon'}\int_0^1 dx \int\frac{ d\Omega'}{(\frac{Q}{(\b p-\b k)^2+\b p^{'2}} x+\frac{\kappa'}{\epsilon'}(1-x))^2}\nonumber\\&=\int\frac{\epsilon^4d\Omega}{(2\pi)^2\kappa((\b p-\b k)^2+\b p^{'2})\epsilon'}\int_0^1 dx \int\frac{ d\Omega'}{1-\frac{b_0^2+i\Gamma^2}{(\b p-\b k)^2+\b p^{'2}} x-\b p'\cdot(\frac{2(\b p-\b k)}{(\b p-\b k)^2+\b p^{'2}} x+\frac{\b k}{\epsilon'\omega}(1-x))}\nonumber\\&=\int\frac{\epsilon^4d\Omega}{(2\pi)\kappa((\b p-\b k)^2+\b p^{'2})\epsilon'}\int_0^1 \frac{ 2dx}{(1-\frac{b_0^2+i\Gamma^2}{(\b p-\b k)^2+\b p^{'2}} x)^2-\b p^{'2}(\frac{2(\b p-\b k)}{(\b p-\b k)^2+\b p^{'2}} x+\frac{\b k}{\epsilon'\omega}(1-x))^2}\nonumber\\&=\epsilon^4\int_{-1}^{1}\frac{2\arctanh(\frac{\abs{\b p'}\sqrt{4\b p^2\kappa^2+(b_0+i\Gamma^2)((b_0^2+i\Gamma^2)-4\epsilon\kappa-2 m^2)}}{(2(\epsilon\epsilon'-m^2)\kappa-\epsilon' (b_0^2+i\Gamma^2))})d\cos{\theta}}{\abs{\b p'}\kappa\sqrt{4\b p^2\kappa^2+(b_0+i\Gamma^2)((b_0^2+i\Gamma^2)-4\epsilon\kappa-2 m^2)}}\,.
\gal
Other integrals can be performed in a similar way and are listed in Appendix~\ref{sec:app3}.
Substituting  \eq{i11},\eq{i12},\eq{i13},\eq{i20},\eq{i27},\eq{i29},\eq{i31} into \eq{imsig1}  we arrive at the final result
\ball{i35}
\frac{d\sigma_\mu}{d\omega}=&\frac{m^2 e^4 \mu ^2  }{6(2\pi)^3 \omega }\frac{ |\b p'|}{|\b p|} \Big[\frac{1}{m^2}(2(\frac{\epsilon'}{\abs{\b p'}}l'+\frac{\epsilon}{\abs{\b p}}l+\frac{\epsilon^2+\epsilon'^2}{\abs{\b p}\abs{\b p'}}ll'-\frac{2m^2}{\abs{\b p}\abs{\b p'}}ll'-1)+\frac{3b_0^2}{\abs{\b p}\abs{\b p'}}ll')\nonumber\\
&+\frac{1}{\Gamma^2}\Re\Big\{\frac{\Gamma^2(5 b_0^4-\Gamma^4)+2b_0^2(2\Gamma^4-b_0^4)i}{m^2\epsilon^4 }I_{1,1}\nonumber\\
&+(3b_0^2\Gamma^2+(\Gamma^4-2b_0^4)i)(\frac{2\b p^2+2\b p'^2}{m^2\epsilon^4 }I_{1,1}-\frac{I_{2,0}+I_{0,2}}{\epsilon^4}+\frac{2\omega}{m^2\epsilon^3}(I_{1,0}-I_{0,1}))\nonumber\\&+\frac{\Gamma^2-2b_0^2i}{\epsilon^4}(-4(\b p^2I_{2,0}-2(\epsilon\epsilon'-m^2)I_{1,1}+\b p'^2I_{0,2})-8\epsilon\omega(I_{1,0}-I_{0,1})+\frac{2\epsilon^2\omega^2}{m^2}(I_{1,-1}+I_{-1,1})) \Big\}\Big]\,
\gal
where $l=2\arctanh(\frac{\abs{\b p}}{\epsilon})$ and $l'=2\arctanh(\frac{\abs{\b p'}}{\epsilon'})$.

\subsection{Ultrarelativistic limit}\label{sec:j}

In practical applications the ultrarelativistic limit $\epsilon,\epsilon'\gg m$ is of most interest. The momentum transfer in this case tends to be small with the minimum occurring for collinear $\b p$, $\b p'$ and $\b k$:
\ball{j-1}
\b q^2_\text{min}=\left(\sqrt{\epsilon'^2-m^2}-\sqrt{\epsilon^2-m^2}+\omega\right)^2\approx \frac{\omega^2 m^4}{4\epsilon^2\epsilon'^2}\,.
\gal
Noting that the resonance contributes to the photon spectrum only when $\b q^2_\text{min}\le b_0^2$ (see e.g.\ \eq{d6}), one finds that the maximum possible photon energy in the resonant/anomalous part of the spectrum is  
\ball{j-3}
\omega_0=\frac{2\epsilon^2b_0}{2\epsilon b_0+m^2}\,.
\gal
One can see this cutoff emerging directly in the integrals \eq{i11},\eq{i12}-\eq{i31}. The anomalous part stems from the imaginary part of those integrals and can be traced back to the kinematic region where the argument of the hyperbolic arctangent exceeds unity. In the ultrarelativistic limit this can occur in the region where $\omega\leq \omega_0$. The explicit expressions for the integrals can be found in Appendix~\ref{sec:app3}.  

In the ultra-relativistic limit \eq{i35} reduces to
\ball{j1}
\frac{d\sigma_\mu}{d\omega}\approx&\frac{ e^4 \mu ^2  }{6(2\pi)^3 \omega }\frac{ \epsilon'}{\epsilon} \Big[2(\ln(\frac{4\epsilon'^2\epsilon^2}{m^4})+(\frac{\epsilon}{\epsilon'}+\frac{\epsilon'}{\epsilon}) \ln(\frac{4\epsilon^2}{m^2})\ln(\frac{4\epsilon'^2}{m^2})-1)\nonumber\\
&+\frac{m^2}{\Gamma^2}\Re\Big\{(3b_0^2\Gamma^2+(\Gamma^4-2b_0^4)i)(\frac{2\epsilon^2+2\epsilon'^2}{m^2\epsilon^4 }I_{1,1})\nonumber\\&+\frac{\Gamma^2-2b_0^2i}{\epsilon^4}(-4(\epsilon^2I_{2,0}-2(\epsilon\epsilon')I_{1,1}+\epsilon'^2I_{0,2})+\frac{2\epsilon^2\omega^2}{m^2}(I_{1,-1}+I_{-1,1})) \Big\}\Big]\,.
\gal
Plugging the individual integrals \eq{j11}--\eq{j19} into \eq{j1} yields
\ball{j3}
\frac{d\sigma_\mu}{d\omega}\approx&\frac{ e^4 \mu ^2  }{3(2\pi)^3 \omega }\frac{ \epsilon'}{\epsilon} \Bigg\{\left(\frac{(3b_0^2+2m^2)(\epsilon^2+\epsilon^{'2})}{2m^2\epsilon'\epsilon }-\frac{\epsilon^2}{ \epsilon^{'2}}-\frac{\epsilon^{'2}}{ \epsilon^{2}}\right)\ln \left(\frac{16 \epsilon^4 \epsilon^{'4}}{m^4 \omega ^4}\right)+4(\frac{\epsilon^2}{ \epsilon^{'2}}  +\frac{\epsilon'^2}{\epsilon^2 })-1\nonumber\\
&+(1-2\frac{\epsilon^2}{ \epsilon^{'2}}) \ln \left(\frac{4 \epsilon^{'2}}{m^2 }\right)+(1-2\frac{\epsilon^{'2}}{ \epsilon^{2}}) \ln \left(\frac{4 \epsilon^{2}}{m^2 }\right)+\left(\frac{\epsilon}{\epsilon'}+\frac{\epsilon'}{\epsilon}\right) \ln(\frac{4\epsilon^2}{m^2})\ln(\frac{4\epsilon^{'2}}{m^2})
  \nonumber\\
&+\frac{b_0^2m^2\pi}{\Gamma^2}\Bigg[\left(\frac{4}{\epsilon}+\frac{(\epsilon^2+\epsilon^{'2})b_0^2}{m^2\epsilon^2\epsilon' }\right)\left(\frac{2\epsilon}{m^2}-\frac{\omega}{b_0 \epsilon'}\right)-4\frac{\epsilon^{'2}}{\epsilon^3}  \left(\frac{\epsilon }{m^2}-\frac{\omega }{2b_0 \epsilon' }\right)\nonumber\\
&-2\frac{\epsilon^2  }{ \epsilon^{'3}}\left(\frac{2\epsilon '}{m^2}-\frac{\omega }{b_0 \epsilon }\right)+\frac{\omega^2}{m^2\epsilon^2}\left(\frac{b_0(\epsilon^2+\epsilon'^2)}{\epsilon^{'2}\omega}-\frac{m^2\epsilon}{2\epsilon'^3}-\frac{m^2}{2\epsilon'\epsilon}\right)\Bigg] \Theta(\omega_0-\omega) \Bigg\}\,,
\gal
where $\Theta$ is the step function. In the limit $b_0\to 0$ this equation reduces to the result  obtained in \cite{Gluckstern:1953zz}.

\begin{figure}[ht]
\begin{tabular}{cc}
      \includegraphics[height=5.5cm]{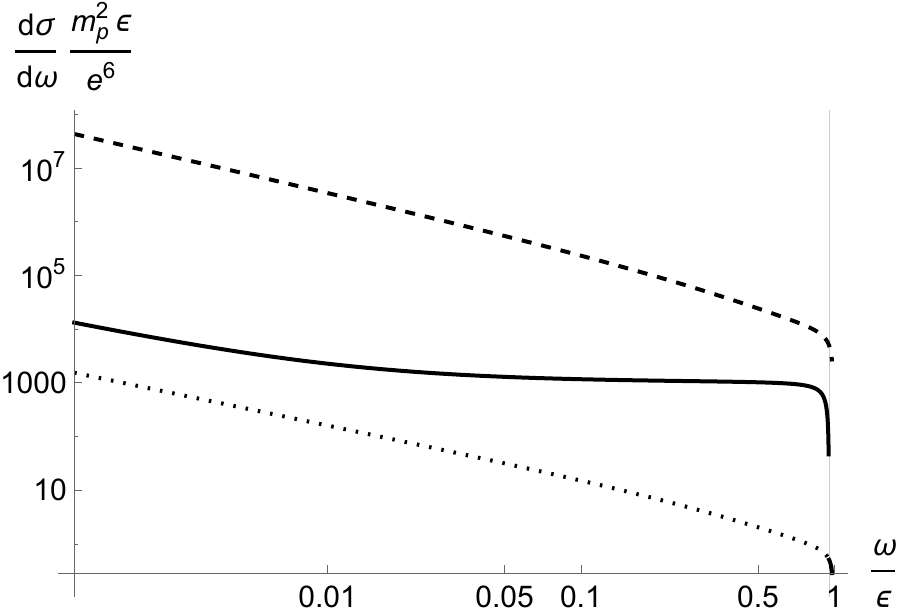} &
      \includegraphics[height=5.5cm]{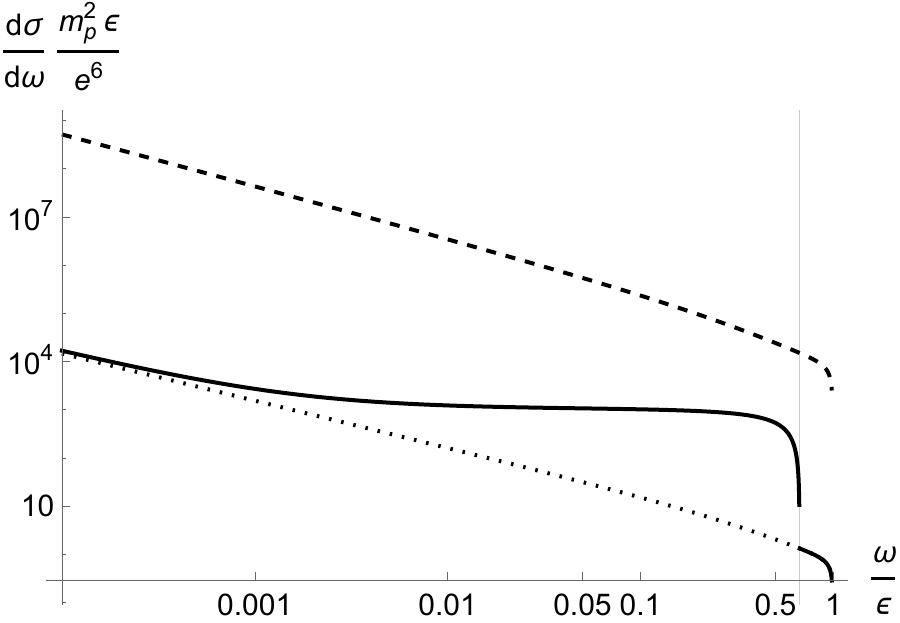}\\
      \includegraphics[height=5.5cm]{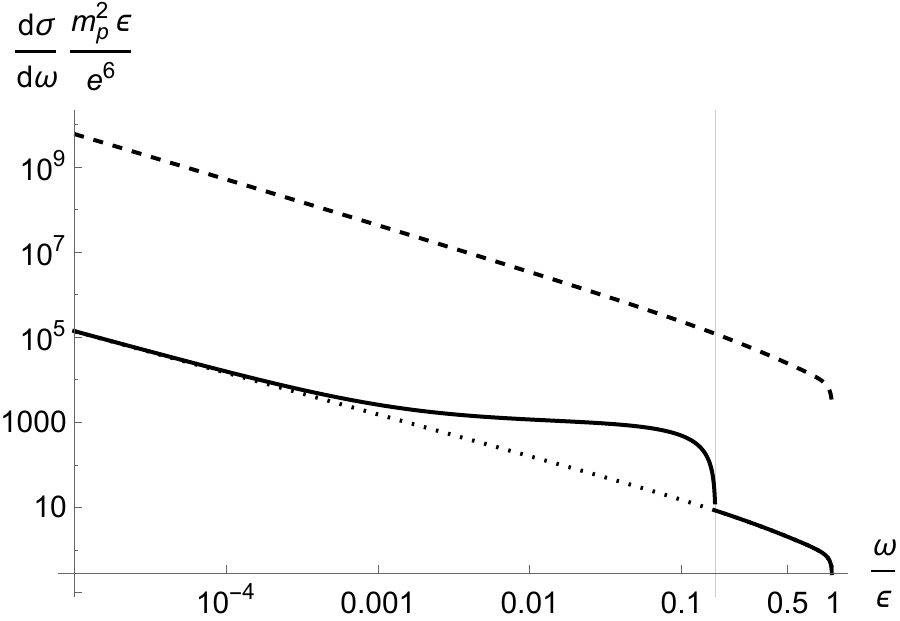} &
      \includegraphics[height=5.5cm]{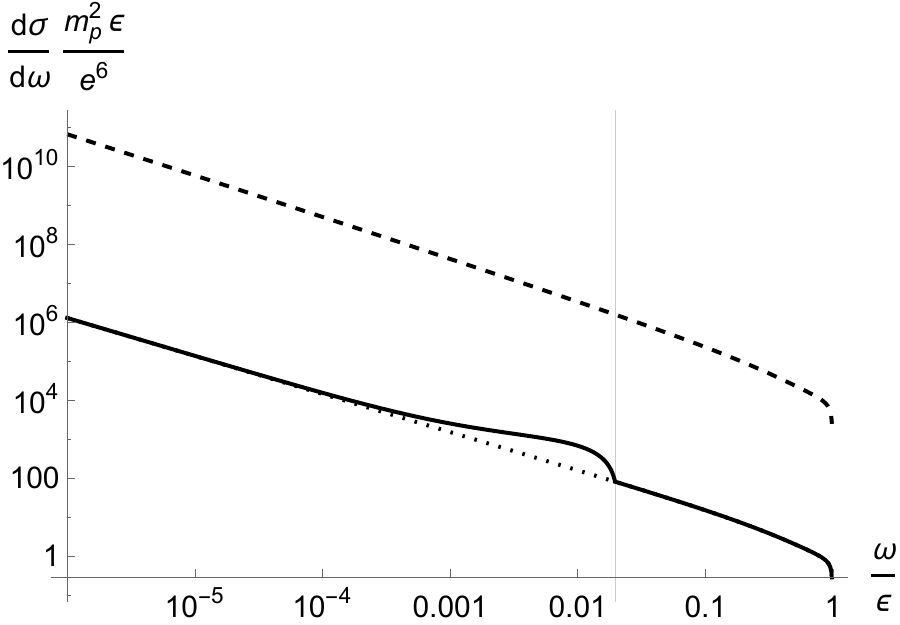}
      \end{tabular}
  \caption{Coulomb (dashed line), the magnetic moment (solid line) and the magnetic moment for $b_0=0$(dotted line) terms of the photon bremsstrahlung cross section in the ultra-relativistic limit  \eq{j3}, for $m=m_p$, $\Gamma=0.01 b_0$, $\epsilon=100 m$, $Z=33$, $\mu =5\mu_N$. From top left to bottom right $b_0=0.1 m,0.01 m,0.001$ and $0.0001 m$ respectively. The Coulomb term is taken from \cite{Berestetskii:1982qgu} in the ultrarelativistic approximation (which reduces to \eq{k5} at $\omega\ll \epsilon$).}
\label{fig:spectrum}
\end{figure}

In the semi-soft photon limit $b_0\ll \omega\ll \epsilon$ the spectrum further simplifies:
\ball{j5}
\frac{d\sigma_\mu}{d\omega}\approx\frac{ 2e^4 \mu ^2  }{3(2\pi)^3 \omega } \left[\frac{3b_0^2}{m^2}\ln(\frac{4\epsilon^4}{m^2\omega^2})+\ln^2\frac{4\epsilon^2}{m^2}+\frac{2b_0^4\pi}{m^2\Gamma^2} \Theta(\omega_0-\omega) \right]\,.
\gal

\section{Discussion}\label{sec:dis}

The main results of this paper are Eqs.~\eq{imsig1},\eq{i35} displaying the contribution of the anomalous chiral magnetic current $\b j= b_0 \b B$ to the differential cross section of the photon bremsstrahlung with energies $\omega\gg b_0$. As we explained in Introduction, the exclusion of the region $\omega\lesssim b_0$ allows us to disregard the chiral Cherenkov radiation which we intend to address in a separate work. In the ultrarelativistic limit the cross section is given by \eq{j3}. 

The two salient features of the anomalous contribution is the enhancement by the large factor $(b_0/\Gamma)^2$  and the emergence of the ultra-violet cutoff \eq{j-3} due to the resonance in the $t$-channel. The order of magnitude of the anomalous contribution can be estimated using \eq{j5}. If the mass of the projectile charge particle is of the same order of magnitude as $b_0$ or smaller, then the anomalous contribution is strongly enhanced. In the opposite limit, 
 $b_0\ll m$, the ratio of the anomalous and the conventional contributions at $\omega<\omega_0$ is of the order $(b_0/\Gamma)^2(b_0/m)^2$ and no general statement can be made. 
 
The total bremsstrahlung spectrum is a sum of the Coulomb $\sigma_e$ and the magnetic moment $\sigma_ \mu$ terms. The Coulomb contribution to the photon emission is given by the amplitude $\mathcal{M}_e$ in \eq{c7}. The corresponding cross section reads in the ultrarelativistic limit for soft photons
\cite{Bethe:1934za,Berestetskii:1982qgu}
\ball{k5}
    \frac{d\sigma_e}{d\omega}\approx \frac{ Z^2e^6}{12(2\pi)^3m^2\omega}\left(\ln\frac{2\epsilon^2}{m\omega}-\frac{1}{2}\right)\,.
\gal
Naturally, the magnetic moment term is suppressed compared to the Coulomb term by a factor $(m/m_p)^2(\mu/\mu_N)^2Z^{-2}$. Here $\mu_N=e/2m_p$ is the nuclear magneton and typical values of the nuclear magnetic moments are in the range $|\mu|=(0.1-10)\mu_N$\cite{Stone:2016bmk}. Considering that the two formulas \eq{j5} and \eq{k5} have similar dependence on photon and projectile energies the Coulomb term overshadows the conventional (i.e.\ $b_0=0$) magnetic moment term. For systems with $b_0\sim m$, the anomalous contribution is enhanced by  $(b_0/\Gamma)^2$ and therefore it can in principle compete with the Coulomb term. Moreover, due to the knee at $\omega=\omega_0$ the anomalous contribution can be clearly distinguished, see \fig{fig:spectrum}.

There are several physical systems where the bremsstrahlung is modified by the chiral anomaly. In the quark-gluon plasma produced in relativistic heavy-ion collisions, it is expected that $b_0$ is induced by the fluctuating topological charge density of the color fields, though its magnitude is quite uncertain \cite{Kharzeev:2013ffa}. Perhaps one can expect that the finite temperature contributions to the photon propagator are about the same order of magnitude as $b_0$ and hence play a significant phenomenological role. Another system with finite $b_0$ is the hypothetic cosmic axion \cite{Sikivie:2020zpn}. In this case  $b_0$ is proportional to the time derivative of the axion field, and thus to the axion mass. Our model describes interaction of the cosmic rays with the axion field. Finally, chiral excitations in Dirac and Weyl semimetals are described by the Hamiltonian with finite $\b b$ \cite{Zyuzin:2012tv,Klinkhamer:2004hg}. Extension of our formalism to anisotropic systems is straightforward though quite laborious (see e.g.\ \cite{Tuchin:2018mte}).

  At any rate, in the physical systems that we are aware of, such as the Weyl semimetals and the axion $b_0\ll m$ implying that anomalous modifications of the photon propagator have small effect on the photon emission. We also expect that similar considerations hold for the gluon emission by a fast quark in the quark-gluon plasma produced in heavy-ion collisions; there the anomaly parameter also seems to be small.  This leaves the chiral Cherenkov effect as the leading source of the anomalous photons in the chiral systems. Nevertheless, we note that the values of the anomaly parameters $b_0$ and $\b b$ can be enhanced in external electric and magnetic fields, see e.g.\cite{Li:2014bha}, making  the anomalous resonance photon production discussed in this paper  phenomenologically relevant.

\acknowledgments
This work  was supported in part by the U.S. Department of Energy under Grant No.\ DE-FG02-87ER40371.

\appendix 
\section{Photon propagator in the chiral medium}\label{sec:app1}

The classical equation of motion of the photon field is 
\ball{ap1}
\left[ g^{\mu\nu}\partial^2-\partial^\mu\partial^\nu-\epsilon^{\mu\nu\alpha\beta}b_\alpha\partial_\beta\right] A_\nu= j^\mu\,.
\gal
It is invariant under the gauge transform 
\ball{ap2}
A_\nu\to A_\nu +\partial_\nu\chi\,
\gal
where $\chi$ is an arbitrary function. In the Lorenz gauge $\partial\cdot A=0$ the photon Green's function obeys the equation
\ball{ap3}
\left[ g^{\mu\nu}\partial^2-(1-1/\xi)\partial^\mu\partial^\nu-\epsilon^{\mu\nu\alpha\beta}b_\alpha\partial_\beta\right] D_{\nu\lambda}(x)= i\delta\indices{^\mu_\lambda} \delta^4(x)\,,
\gal
where $\xi$ is the gauge parameter. In momentum space \eq{ap3} becomes
\ball{ap5}
\left[ -k^2g^{\mu\nu}-(1-1/\xi)k^\mu k^\nu+i\epsilon^{\mu\nu\alpha\beta}b_\alpha k_\beta\right] D_{\nu\lambda}(k)= i\delta\indices{^\mu_\lambda}\,,
\gal
which is solved by
\ball{ap7}
 D^{\nu\lambda}(k)= &-\frac{i}{k^4+b^2k^2-(k\cdot b)^2}
\bigg\{ k^2 g^{\nu\lambda}+b^\nu b^\lambda+i\epsilon_{\nu\lambda\alpha\beta}b^\alpha k^\beta
\nonumber\\
&- \frac{(b\cdot k)}{k^2}(k^\nu b^\lambda+k^\lambda b^\nu)
+\left[b^2\xi-(1-\xi)\left( k^2-\frac{(b\cdot k)^2}{k^2}\right)\right]\frac{k^\lambda k^\nu}{k^2}
\bigg\}
\gal
The terms in the second line vanish when $D^{\nu\lambda}$ is inserted in the Feynman diagram due to the current conservation. The terms in the first line yield \eq{a5}. 

A more general expression for the propagator can be obtained in a covariant gauge $(\partial+C b)\cdot A=0$, where $C$ is a number. 

\section{Photon polarization sum}\label{sec:app2}

The wavefunction of a photon at $b=(b_0,\b 0)$ is a plane wave (check normalization)
\ball{ap20}
A^\mu(x) =\frac{1}{\sqrt{2\omega}}e^\mu_k e^{-ik\cdot x}\,,
\gal
where $e^\mu_k= (0,\b e_k)$ is a circular polarization vector satisfying the 
 the Lorenz gauge $k\cdot e_k=\b k\cdot \b e_k=0$. Once the wave functions \eq{ap20} are chosen, there is no remaining gauge freedom. Only circularly polarized photons solve the equation of motion \eq{ap1}. 
 
 This is in sharp contrast with the free photon wavefunction which is also given by the plane wave as in \eq{ap20}. However, since $k^2=0$, there still remains gauge freedom to transform the polarization vector  $e^\mu\to e^\mu +\chi k^\mu$ without changing the form of the wavefunction or violating the Lorenz gauge condition. 
 
The photon polarization sum is
\ball{ap23}
d_k^{ij}= \sum_\text{pol}e^i_ke^{*j}_k=\delta^{ij}-\frac{k^ik^j}{\b k^2}\,.
\gal 
There are no remaining gauge freedom to transform it to any other form. In particular $d^{ij}$ cannot be replaced with $-g^{\mu\nu}$. As an illustration consider the amplitude $e_k\cdot \mathcal{M}$. Let $k$ be in the $z$-direction, then using the current conservation $k\cdot \mathcal{M}=0$ we can write
$$
\sum_\text{pol}|e_k\cdot \mathcal{M}|^2=|\mathcal{M}_x|^2+|\mathcal{M}_y|^2= |\mathcal{M}_x|^2+|\mathcal{M}_y|^2+|\mathcal{M}_z|^2-\frac{\omega^2}{\b k^2}|\mathcal{M}_0|^2\neq -g^{\mu\nu}\mathcal{M}_\mu\mathcal{M}^*_\nu\
$$
since $k^2= \omega^2-\b k^2\neq 0$. 

Introducing the unit vector $n^\mu=b^\mu/b_0= (1,\b 0)$ we can cast the polarization sum \eq{ap23} in the boost-invariant form:
\ball{ap28}
d^{\mu\nu}= -g^{\mu\nu}-\frac{k^\mu k^\nu-(k\cdot n)(k^\mu n^\nu+k^\nu n^\mu)+n^\mu n^\nu k^2}{(k\cdot n)^2-k^2}\,.
\gal
Its spatial components reduce to \eq{ap23}, while the other ones $d^{00}=d^{0i}=0$. 

\section{Angular integrals}\label{sec:app3}

\ball{i12}
    I_{-1,1}=\epsilon^2\int \frac{\kappa d\Omega d\Omega'}{(2\pi)^2\kappa'Q}=\epsilon^2\int_{-1}^{1}\frac{2\kappa*\arctanh(\frac{\abs{\b p'}\sqrt{4\b p^2\kappa^2+(b_0+i\Gamma^2)((b_0^2+i\Gamma^2)-4\epsilon\kappa-2 m^2)}}{(2(\epsilon\epsilon'-m^2)\kappa-\epsilon' (b_0^2+i\Gamma^2))})d\cos{\theta}}{\abs{\b p'}\sqrt{4\b p^2\kappa^2+(b_0+i\Gamma^2)((b_0^2+i\Gamma^2)-4\epsilon\kappa-2 m^2)}}
\gal
\ball{i13}
    I_{0,1}=\epsilon^3\int \frac{ d\Omega d\Omega'}{(2\pi)^2\kappa'Q}=\epsilon^3\int_{-1}^{1}\frac{2\arctanh(\frac{\abs{\b p'}\sqrt{4\b p^2\kappa^2+(b_0+i\Gamma^2)((b_0^2+i\Gamma^2)-4\epsilon\kappa-2 m^2)}}{(2(\epsilon\epsilon'-m^2)\kappa-\epsilon' (b_0^2+i\Gamma^2))})d\cos{\theta}}{\abs{\b p'}\sqrt{4\b p^2\kappa^2+(b_0+i\Gamma^2)((b_0^2+i\Gamma^2)-4\epsilon\kappa-2 m^2)}}
\gal
\ball{i20}
    I_{2,0}=\epsilon^4\int \frac{d\Omega d\Omega'}{(2\pi)^2\kappa^2Q}=\epsilon^4\int \frac{d\Omega }{(2\pi)^2\kappa^2}\int\frac{ d\Omega'}{Q}=\epsilon^4\int_{-1}^{1}\frac{\arctanh(\frac{2\abs{\b p'}\sqrt{p^{'2}+2\omega\kappa}}{2p^{'2}+2\omega\kappa-b_0^2-i\Gamma^2})d\cos{\theta}}{\abs{\b p'}\kappa^2\sqrt{p^{'2}+2\omega\kappa}}
\gal
The rest of the integrals can be related to Eqs.~\eq{i11} and \eq{i12}--\eq{i20} by means of the transformation $\b p\to \b p'$, $\b p'\to \b p$, $\omega \to -\omega$ and therefore  
\ball{i23}
    \b q^2=(\b p'-\b p+\b k)^2\rightarrow(\b p-\b p'-\b k)^2=(\b p'-\b p+\b k)^2=\b q^2\,,
\gal
\ball{i25}
    \kappa= \epsilon-\frac{\b p \cdot \b k}{\omega}\rightarrow \epsilon'-\frac{\b p' \cdot \b k}{\omega}=\kappa'\,.
\gal
Taking advantage of these transformations we derive
\ball{i27}
    &I_{0,2}=\epsilon^4\int \frac{d\Omega d\Omega'}{(2\pi)^2\kappa'^2Q}=\epsilon^4\int_{-1}^{1}\frac{\arctanh(\frac{2\abs{\b p}\sqrt{p^2-2\omega\kappa'}}{2\b p^2-2\omega\kappa'-b_0^2-i\Gamma^2})d\cos{\theta'}}{\abs{\b p}\kappa'^2\sqrt{p^2-2\omega\kappa'}}
\gal
\ball{i29}
    I_{1,-1}&=\epsilon^2\int \frac{\kappa'd\Omega d\Omega'}{(2\pi)^2\kappa Q}=\epsilon^2\int_{-1}^{1}\frac{2\kappa'*\arctanh(\frac{\abs{\b p}\sqrt{4\b p^{'2}\kappa^{'2}+(b_0+i\Gamma^2)((b_0^2+i\Gamma^2)-4\epsilon'\kappa'-2 m^2)}}{(2(\epsilon\epsilon'-m^2)\kappa'-\epsilon (b_0^2+i\Gamma^2))})d\cos{\theta'}}{\abs{\b p}\sqrt{4\b p^{'2}\kappa^{'2}+(b_0+i\Gamma^2)((b_0^2+i\Gamma^2)-4\epsilon'\kappa'-2 m^2)}}
\gal
\ball{i31}
    I_{1,0}&=\epsilon^3\int \frac{d\Omega d\Omega'}{(2\pi)^2\kappa Q}=\epsilon^3\int_{-1}^{1}\frac{2\arctanh(\frac{\abs{\b p}\sqrt{4\b p^{'2}\kappa^{'2}+(b_0+i\Gamma^2)((b_0^2+i\Gamma^2)-4\epsilon'\kappa'-2 m^2)}}{(2(\epsilon\epsilon'-m^2)\kappa'-\epsilon (b_0^2+i\Gamma^2))})d\cos{\theta'}}{\abs{\b p}\sqrt{4\b p^{'2}\kappa^{'2}+(b_0+i\Gamma^2)((b_0^2+i\Gamma^2)-4\epsilon'\kappa'-2 m^2)}}
\gal

In the ultrarelativistic limit $\epsilon\gg m$ the remaining integral can be done explicitly with the following result:
\ball{j11}
    I_{1,1}\approx\frac{\epsilon^3\ln(\frac{4\epsilon^2\epsilon'^2}{m^2\omega^2})}{m^2\epsilon'}+i\frac{\pi\epsilon^2}{2\epsilon'}(\frac{2\epsilon}{m^2}-\frac{\omega}{b_0 \epsilon'})\Theta(\omega_0-\omega)\,,
\gal
\ball{j13}
    I_{1,-1}+I_{-1,1}\approx\frac{2\epsilon\ln(\frac{4\epsilon^2\epsilon'^2}{m^2\omega^2})}{\epsilon'}+i\pi\left(\frac{b_0(\epsilon^2+\epsilon'^2)}{2\epsilon'^2\omega}-\frac{m^2\epsilon}{4\epsilon'^3}-\frac{m^2}{4\epsilon'\epsilon}\right) \Theta(\omega_0-\omega)\,,
\gal
\ball{j15}
    I_{1,0}-I_{0,1}\approx\left(\frac{\epsilon^2\ln(\frac{4\epsilon'^2}{m^2})}{2\epsilon'^2}-\frac{\epsilon\ln(\frac{4\epsilon^2}{m^2})}{2\epsilon'}\right)\ln(\frac{4\epsilon^2\epsilon'^2}{m^2\omega^2})+\frac{i\pi\epsilon\omega\ln(\frac{2b_0\epsilon\epsilon'}{m^2\omega})}{2\epsilon'^2}\Theta(\omega_0-\omega)\,,
\gal
\ball{j17}
    I_{2,0}\approx\frac{\epsilon^4  \left(\ln \left(\frac{16 \epsilon^2 \epsilon^{'4}}{m^4 \omega ^2}\right)-2\right)}{m^2 \epsilon ^{'2}} + \frac{i \pi\epsilon^4  \left(\frac{2\epsilon '}{m^2}-\frac{\omega }{b_0 \epsilon }\right)}{2 \epsilon'^3} \Theta(\omega_0-\omega)\,,
\gal
\ball{j19}
    I_{0,2}\approx\frac{\epsilon^2  \left(\ln \left(\frac{16 \epsilon^4 \epsilon'^2}{m^4 \omega ^2}\right)-2\right)}{\text{m}^2 } +  i \pi\epsilon  \left(\frac{\epsilon }{m^2}-\frac{\omega }{2b_0 \epsilon' }\right) \Theta(\omega_0-\omega)\,.
\gal


\end{document}